\documentclass[runningheads]{llncs}

\usepackage{times}
\usepackage{graphicx}
\usepackage[square,comma,numbers,sort&compress,sectionbib]{natbib}
\usepackage{latexsym}
\usepackage{comment}
\usepackage{cleveref}
\usepackage{url}
\usepackage{multirow}

\usepackage{booktabs}
\usepackage{adjustbox}
\usepackage[subtle]{savetrees}
\usepackage{wrapfig}
\usepackage{floatrow}
\usepackage{paralist}

\input{addon/commands}
\graphicspath{{figures/}}

\begin{document}

\title{QRFA:~A~Data-Driven~Model~of Information-Seeking Dialogues}

\author{%
 	Svitlana Vakulenko\inst{1}
 	\and
 	Kate Revoredo\inst{2}
 	\and
 	Claudio Di Ciccio\inst{1}
 	\and
 	Maarten de Rijke\inst{3}}

\institute{Vienna University of Economics and Business, Vienna, Austria\\
\email{\{svitlana.vakulenko, claudio.di.ciccio\}@wu.ac.at}
\and
Graduate Program in Informatics, Federal University of Rio de Janeiro, Rio de Janeiro, Brazil\\
\email{katerevoredo@ppgi.ufrj.br}
\and
University of Amsterdam, Amsterdam, The Netherlands\\
\email{derijke@uva.nl}}

\maketitle

\begin{abstract}
Understanding the structure of interaction processes helps us to improve information-seeking dialogue systems. 
Analyzing an interaction process boils down to discovering patterns in sequences of alternating utterances exchanged between a user and an agent. 
Process mining techniques have been successfully applied to analyze structured event logs, discovering the underlying process models or evaluating whether the observed behavior is in conformance with the known process. 
In this paper, we apply process mining techniques to discover patterns in conversational transcripts and extract a new model of information-seeking dialogues, QRFA, for Query, Request, Feedback, Answer.
Our results are grounded in an empirical evaluation across multiple conversational datasets from different domains, which was never attempted before.
We show that the QRFA model better reflects conversation flows observed in real informa\-tion-seeking conversations than models proposed previously. 
Moreover, QRFA allows us to identify malfunctioning in dialogue system transcripts as deviations from the expected conversation flow described by the model via conformance analysis.

\keywords{Conversational search \and Log analysis \and Process mining.}
\end{abstract}


\section{Introduction}
\label{sec:intro}

Interest in information-seeking dialogue systems is growing rapidly, in information retrieval, language technology, and machine learning.
There is, however, a lack of theoretical understanding of the functionality such systems should provide~\cite{trippas2018informing}.
Different information-seeking models of dialogue systems use different terminology as well as different modeling conventions, and conversational datasets are annotated using different annotation schemes~\cite[see, e.g.,][]{trippas2016how,williams2016dialog}.
These discrepancies hinder direct comparisons and aggregation of the results.
Moreover, the evaluation of conversational datasets has largely been conducted based on manual efforts.
Clearly, it is infeasible to validate models on large datasets without automated techniques.

Against this background, we create a new annotation framework that is able to generalize across conversational use cases and bridge the terminology gap between diverse theoretical models and annotation schemes of the conversational datasets collected to date. 
We develop and evaluate a new information-seeking model, which we name QRFA, for Query, Request, Feedback, Answer, which shows better performance in comparison with previously proposed models and helps to detect malfunctions from dialogue system transcripts.
It is based on the analysis of 15,931 information-seeking dialogues and evaluated on the task of interaction success prediction in 2,118 held-out dialogues.

The QRFA model is derived and evaluated using process mining techniques~\cite{Aalst16}, which makes this approach scalable. 
We view every conversation to be an instance of a general information-seeking process. 
This inclusive perspective helps us to extract and generalize conversation flows across conversations from different domains, such as bus schedules and dataset search.

To the best of our knowledge, we present the first grounded theory of information-seeking dialogues that is empirically derived from a variety of conversational datasets.
Moreover, we describe the methodology we used to develop this theory that can be used to revise and further extend the proposed theory.
We envision that the model and the approach that we describe in this paper will help not only to better understand the structure of information-seeking dialogues but also to inform the design of conversational search systems, their evaluation frameworks and conversational data sampling strategies.
More concretely, we discovered a set of functional components for a conversational system as different interaction patterns and the distribution over the space of next possible actions. 

The remainder of the paper is organized as follows. 
Section~\ref{sec:background} provides a short review of discourse analysis and a gentle introduction to process mining. 
In Section~\ref{sec:related_work} we discuss related work in two contexts: theoretical models of conversational search and process mining for discourse analysis. 
Section~\ref{sec:approach} provides details of our approach to mining processes from conversations. 
In Section~\ref{sec:experiments} we report on the results of applying our conversation mining approach to several conversational datasets and we describe the model we obtained as a result. 
We conclude in Section~\ref{sec:conclusion}.


\section{Background}
\label{sec:background}

\subsection{Discourse analysis}

A plethora of approaches have been proposed for discourse analysis, all focusing on different aspects of communication processes~\cite{schiffrin1994approaches}. 
The main bottleneck in traditional approaches to discourse analysis, especially ones grounded in social and psychological theories \cite[e.g.,][]{goffman1988erving}, is context-dependence of the conversational semantics. 
Many studies rely on a handful of sampled or even artificially constructed conversations to illustrate and advocate their discourse theories, which limits their potential for generalization.

In this work we describe and demonstrate the application of a data-driven approach that can be applied to a large volume of conversational data, to identify patterns in the conversation dynamics. 
It is directly rooted in Conversational Analysis (CA)~\cite{schegloff1968sequencing}, which proposes to analyze regularities such as adjacency pairs and turn-taking in conversational structures, and Speech Act Theory (SAT)~\cite{searle1969speech,austin1976things} to identify utterances with functions enabled through language (speech acts). 
To this end we leverage state-of-the-art techniques developed in the context of process mining~\cite{Aalst16}, which has traditionally been applied in the context of operational business processes such as logistics and manufacturing, to discover and analyze patterns in sequential data. 

Process mining (PM) has been designed to deal with structured data organized into a process log rather than natural language, such as conversational transcripts. 
However, we view a conversation as a sequence of alternating events between a user and an agent, thus a special type of process, a communication or information exchange process, that can be analyzed using PM by converting conversational transcripts into process logs. 
Basic concepts and techniques from PM, which we adopt in our discourse analysis approach, are described in the next subsection. 

\subsection{Process mining}

A process is a structure composed of events aligned between each other in time. 
The focus of PM is on extracting and analyzing process models from event logs. 
Each event in the log refers to the execution of an activity in a process instance. 
Additional information such as a reference to a resource (person or device) executing the activity, a time stamp of the event, or data recorded for the event, may be available.

Two major tasks in PM are \emph{process discovery} and \emph{conformance checking}. 
The former is used to extract a process model from an event log, and the latter to verify the model against the event log, i.e., whether the patterns evident from the event log correspond to the structure imposed by the model. 
It is possible to verify conformance against an extracted model as well as against a theoretical (independently constructed) model.

In this work, we adopt state-of-the-art PM techniques to analyze conversational transcripts by extracting process models from publicly available datasets of information-seeking dialogues, and to verify and further extend a theoretical model of information-seeking dialogues based on the empirical evidence from these corpora.


\section{Related Work}
\label{sec:related_work}

\subsection{Theoretical models of information-seeking dialogues}

\begin{figure}[!t]
\centering
\includegraphics[width=0.65\columnwidth]{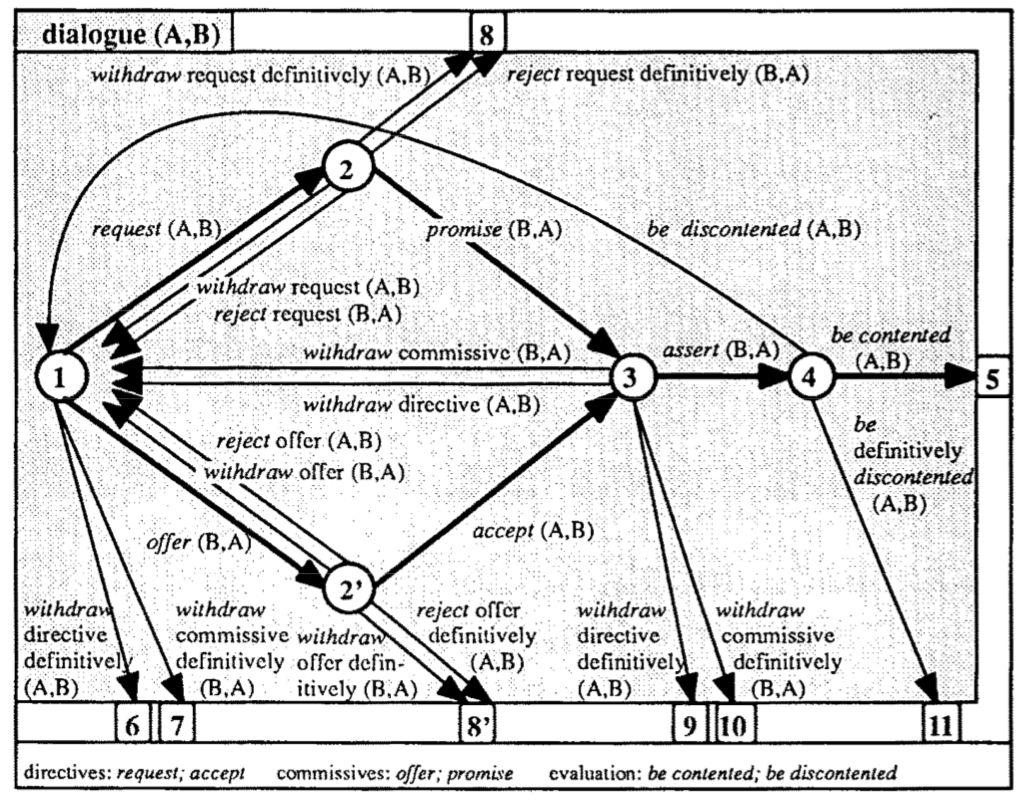}
\caption{COnversational Roles (COR) model of information-seeking dialogues proposed by \citet{sitter1992modeling}.}
\label{fig:model_sitter}
\end{figure}

The first theoretical model of information-seeking dialogues has been proposed by \citet{winograd1986understanding} and  further extended by \citet{sitter1992modeling} to the COnversational Roles (COR) model.
The authors envision an implementation of a human-computer dialogue system that could support necessary functionality to provide efficient information access and illustrate it as a transition network over a set of speech acts (Figure~\ref{fig:model_sitter}).
This model describes a use case of a ``conversation for action'' and is mainly focused on tracking commitments rather than analyzing language variations.
We use the COR model as our baseline and show in an empirical evaluation that it is not able to adequately reflect the structure of information-seeking dialogues across four publicly available datasets and propose an alternative model.

\citet{belkin1995cases} argue for a modular structure of an interactive information retrieval (IR) system that would be able to support various dialogue interactions. 
The system should be able to compose interactions using a set of scripts, which provide for various information-seeking strategies (ISSs) that can be described using the COR model. 
The authors introduce four dimensions to describe different ISSs and propose to collect cases for each of the ISSs to design the scripts. 
We closely follow their line of work by accumulating empirical evidence from publicly available conversational datasets to validate both the COR model and the ISS dimensions proposed by \citet{sitter1992modeling} that form the basis for accumulating the body of sample scripts describing various ISSs.

\begin{figure}[!t]
\centering
\vspace*{-5mm}
\includegraphics[width=0.9\textwidth]{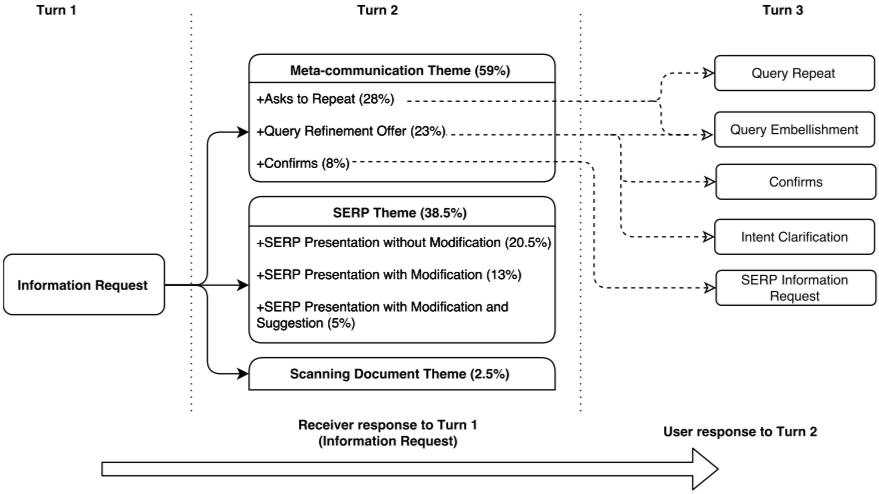}
\caption{Spoken Conversational Search (SCS) model by \citet{trippas2016how} (SERP is short for Search Engine Results Page.)}
\label{fig:model_trippas}
\end{figure}

More recently, \citet{DBLP:conf/chiir/RadlinskiC17} have also proposed a theoretical model for a conversational search system. 
They propose a set of five actions available to the agent and a set of five possible user responses to describe the user-system interactions. 
However, they do not describe in detail the conversation flow between these actions. 
In contrast, \citet{trippas2016how} empirically derive a model of Spoken Conversational Search (SCS) based on collected conversational transcripts (Figure~\ref{fig:model_trippas}).
The SCS model describes the case of using a web search engine via a speech-only interface and is limited to sequences of three conversation turns.
We show how such an empirical approach to analyzing and structuring conversation transcripts into conversation models can be performed at scale on multiple conversational datasets (including SCS) using process mining techniques.

\subsection{Process mining from conversations}

There are relatively few prior publications that demonstrate the benefit of applying process mining (PM) techniques to conversational data.  
\citet{DBLP:journals/internet/CiccioM13} use a corpus of e-mail correspondence to illustrate how the structure of a complex collaborative process can be extracted from message exchanges. 
\citet{wang2015analytical} analyze a sample of discussion threads from an on-line Q\&A forum by applying process mining and network analysis techniques and comparing patterns discovered across different thread categories based on their outcomes (solved, helpful and unhelpful threads).

\citet{DBLP:conf/bpm/RichettiGBS17} analyze the performance of a customer support service team by applying process mining to conversational transcripts that were previously annotated with speech acts using a gazetteer. 
Their results reveal similar structural patterns in the conversational flow of troubleshooting conversations with different durations, i.e., less and more complex cases, which require additional information seeking loops.

To the best of our knowledge, there is no prior work going beyond the individual use cases mining conversations from a specific domain. 
In contrast, we analyze multiple heterogeneous conversational datasets from various domains to be able to draw conclusions on structural similarities as well as differences stemming from variance introduced by labeling approaches and specific characteristics of the underlying communication processes. 

Before applying the proposed approach, utterances have to be annotated with activity labels, such as speech acts~\cite{searle1976classification}. 
The task of utterance classification is orthogonal to our work. 
Dialogue corpora to be used for process mining can be manually annotated by human annotators or automatically by using one of the classification approaches proposed earlier~\cite{stolcke2000dialogue,DBLP:conf/emnlp/CohenCM04,DBLP:conf/emnlp/JeongLL09,DBLP:conf/emnlp/JoYJR17}.


\section{An Empirical Approach to Extracting a Conversation Model from Conversational Transcripts}
\label{sec:approach}


We consider every conversation $C$ in a transcript $\mathcal{C}$ to be an instance of the same communication process, the model of which we aim to discover. 
A conversation is represented as a sequence of utterances $C = \langle u_1, u_2, \ldots\rangle$.
An \textit{utterance}, in this case, is defined rather broadly as a text span within a conversation transcript attributed to one of the conversation participants and explicitly specified during the annotation process (utterance labeling step).
We denote the set of all utterances as $U = \{ u_1, u_2, \ldots\}$. 
Our approach to conversational modeling consists of three steps: (1)~utterance labeling, (2)~model discovery, and (3)~conformance checking, which is useful for model validation and error detection in conversation transcripts.

\subsection{Utterance labeling}

A utterance $u_i$ in a conversation $C$ can be mapped to multiple labels $l^1_i,\ldots,l^n_i$, each belonging to pre-defined label sets $L^1,\ldots,L^n$, respectively.
The label sets, not necessarily disjoint, may correspond to different annotation schemes.
We denote the general multi-labeling of utterances as a mapping function $\hat{\lambda} : U \to L_1 \times \cdots \times L_n$.
It can stem from manual annotations of different human analysts, or categories returned by multiple machine-learned classifiers.
For the sake of readability, we assume in the remainder of this section that all annotations share a single set of labels $L$, thus the mapping function used henceforth is reduced to $\lambda : U \to L$.

\subsection{Extracting the model of the conversation flow}

We apply a process discovery approach to collect patterns of the conversation structure from  transcripts. 
The goal of process discovery is to extract a model that is representative of empirically observed behavior
stored in an \emph{event log}~\cite{Aalst16}.
Event logs can be abstracted as sets of sequences (\emph{traces}), where each element in the sequence (the \emph{event}) is labeled with an activity (\emph{event class}) plus optional \emph{attributes}.
We reduce a conversation transcript $\mathcal{C}$ along with its labeling to an event log, as follows:
a conversation $C$ is a trace, an utterance $u$ is an event, and the label of $u$, $\lambda(u)=l$, is the event class.

There is a wide variety of algorithms that can be applied for process discovery.
Imperative workflow mining algorithms, such as the seminal $\alpha$-algorithm~\cite{Aalst16} or the more recent Inductive Miner (IM)~\cite{DBLP:conf/bpm/LeemansFA14}, extract procedural process models that depict the possible process executions, in the form of, e.g., a Petri net~\cite{Aalst/JCSC1998:ApplicationofPetriNetsToWorkflowManagement}.
Other approaches, such as frequent episode mining~\cite{DBLP:conf/simpda/LeemansA14} or declarative constraints mining~\cite{DiCiccio.Mecella/CIDM2013:TwoStepFast,DiCiccio.Mecella/ACMTMIS2015:DiscoveryDeclarativeControl},
extract local patterns and aggregate relations between activities.
%
%
%
One such relation is the \emph{succession} between two activities, denoting that the second one occurs eventually after the first one. 
In our context, succession between $l_i$ and $l_j$ holds true in a sub-sequence $\langle u_i, \ldots, u_j\rangle$, with $i < j$, if $u_j \mapsto l_j$ and $u_i \mapsto l_i$. 
The frequency of such patterns observed across conversations can indicate dependencies between the utterance labels $l_i$ and $l_j$. Those dependencies can be used to construct a model describing a frequent behavior (model discovery) as well as to detect outliers breaking the expected sequence (error analysis), upon the setting of thresholds for minimum frequency.
The discovery algorithm described in \cite{DiCiccio.Mecella/ACMTMIS2015:DiscoveryDeclarativeControl} requires a linear pass through each sequence to count, for every label $l \in L$,
\begin{inparaenum}[(1)]
	\item its number of occurrences per sequence, and 
	\item the distance at which other labels occur in the same sequence in terms of number of utterances in-between.
\end{inparaenum}

\subsection{Conformance analysis and model validation}

The goal of the model validation step is conformance checking, i.e., to assess to which extent the patterns evident from the transcript fit the structure imposed by the model.
We use it here to also evaluate the predictive power of the model, i.e., the ability of the model to generalize to unseen instances of the conversation process. 
A good model should fit the transcripts but not overfit it.
The model to be validated against can be the one previously extracted from transcripts, or a theoretical model, i.e., an independently constructed one. 
In the former case, we employ standard cross-validation techniques by creating a test split separate from the development set that was used to construct the model.

Model quality can be estimated with respect to a conversation transcript. Likewise, the quality of the conversation can be estimated with respect to a pre-defined model. 
In other words, discrepancies between a conversation model and a transcript indicate either inadequacy of the model or errors (undesired behavior or recording malfunctions) in the conversation. 

To compute fitness, i.e., the ability of the model to replay the event log, we consider the measure first introduced in \cite{AalstAD12WIDM}, based on the concept of alignment.
Alignments keep consistent the replay of the whole sequence and the state of the process by adding so-called non-synchronous moves if needed.
The rationale is, the more non-synchronous moves are needed, the lower the fitness is.
Thus, a penalty is applied by means of a cost function on non-synchronous moves.
For every sequence fitness is computed as the complement to 1 of the total cost of the optimal alignment, divided by the cost of the worst-case alignment.
Log fitness is calculated by averaging the sequence fitness values over all sequences in a log.


\section{QRFA Model Development and Results}
\label{sec:experiments}
In this section, we apply the approach to extraction and validation of a conversational model proposed in Section~\ref{sec:approach} to develop a new model (QRFA). 
We (1)~collect \textit{datasets} of publicly available corpora with information-seeking dialogue transcripts; (2)~analyze and link their annotation schemes to each other and to the COR model (Figure~\ref{fig:model_sitter}); (3)~analyze the conversation flows in the datasets; and (4)~\textit{evaluate} QRFA and compare the results with COR as a baseline.

\subsection{Conversational datasets}
We used publicly available datasets of information-seeking dialogues that are annotated with utterance-level labels (see the dataset statistics in Table~\ref{datasets-table}).

\begin{table}[h]
\vspace*{-5mm}
	\BottomFloatBoxes
	\begin{floatrow}
		\ttabbox{%
			\begin{small}
				\begin{tabular}{l rrrr }
					\toprule
					\bf Dataset & \bf Dialogues & \bf Utterances & \bf Labels &  \\ \midrule
					SCS         & 39            &            101 &         13 &  \\
					ODE         & 26            &            417 &         20 &  \\
					DSTC1       & 15,866        &        732,841 &         37 &  \\
					DSTC2       & 2,118         &         40,854 &         21 &  \\ \bottomrule
				\end{tabular}
			\end{small}
		}{%
			\caption{Dataset statistics.}%
			\label{datasets-table}%
		}
		\hspace*{-.5em}
		\ttabbox{%
			\centering
			\begin{small}
				\begin{tabular}{l @{\hspace{.5em}}ll >{\hspace{1em}}l @{\hspace{.5em}}ll }
					\toprule
					& \multicolumn{2}{c}{Proactive}  &  &  \multicolumn{2}{c}{Reactive}   \\
					\cmidrule{2-3} \cmidrule{5-6}
					\textit{User} & \textbf{Q}uery   & Information &  & \textbf{F}eedback & Positive    \\
					&                  & Prompt      &  &                   & Negative    \\
					\textit{Agent}                              & \textbf{R}equest & Offer       &  & \textbf{A}nswer   & Results     \\
					&                  & Understand  &  &                   & Backchannel \\
					&      \multicolumn{2}{c}{}      &  &                   & Empty       \\ \bottomrule
				\end{tabular} 
			\end{small}%
		}{%
			\caption{New functional annotation schema for information-seeking conversation utterances.} %
			\label{schema-table} %
		}
	\end{floatrow}
\vspace*{-5mm}	
\end{table}

\mysubsubsection{Spoken Conversational Search.} This dataset%
\footnote{\scriptsize\url{https://github.com/JTrippas/Spoken-Conversational-Search}}~\cite[\textit{SCS},][]{trippas2016how} contains human-human conversations collected in a controlled laboratory study with 30 participants.
The task was designed to follow the setup, in which one of the conversation participants takes over the role of the information Seeker and another of the Intermediary between the Seeker and the search engine.
It is the same dataset that was used to develop the SCS model illustrated in Figure~\ref{fig:model_trippas}. 
All dialogues in the dataset are very short and contain at most three turns, with one label per utterance. 
The efficiency of the interaction and the user satisfaction from the interaction are not clear.


\mysubsubsection{Open Data Exploration.} This dataset%
\footnote{\scriptsize\url{https://github.com/svakulenk0/ODExploration_data}} (\textit{ODE}) was collected in a laboratory study with 26 participants and a setup similar to the SCS but with the task formulated in the context of conversational browsing, in which the Seeker does not communicate an explicit information request. 
The goal of the Intermediary is to iteratively introduce and actively engage the Seeker with the content of the information source. 
All dialogues in this dataset contain one label per utterance. 
The majority of the conversation transcripts (92\%) exhibit successful interaction behavior leading to a positive outcome, such as satisfied information need and positive user feedback (only 2 interactions were unsuccessful), and can be considered as samples of effective information-seeking strategies.

\mysubsubsection{Dialog State Tracking Challenge.} These datasets%
\footnote{\scriptsize\url{https://www.microsoft.com/en-us/research/event/dialog-state-tracking-challenge}}~\cite[\textit{DSTC1} and \textit{DSTC2}, ][]{williams2016dialog} provide annotated human-computer dialogue transcripts from an already implemented dialogue system for querying bus schedules and a restaurant database. 
The transcripts may contain more than one label per utterance, which is different from the previous two datasets. 
The efficiency of the interaction and user satisfaction from the interaction with the agent are not clear.

\subsection{QRFA model components}
Since all datasets and the theoretical model that we consider use different annotation schemes, we devise a single schema to be able to aggregate and compare conversation traces.
To the best of our knowledge, no such single schema that is able to unify annotations across a diverse set of information-seeking conversation use cases has been proposed and evaluated before.
Our schema is organized hierarchically into two layers of abstraction to provide a more simple and general as well as more fine-grained views on the conversation components.

First, we separate utterances into four basic classes: two for User (Query and Feedback) and two for Agent (Request and Answer).
This distinction is motivated by the role an utterance plays in a conversation.
Some of the utterances explicitly require a response, such as a question or a request, while others constitute a response to the previous utterance, such as an answer.
Such a distinction is reminiscent of the Forward and Backward Communicative Functions that are foundational for the DAMSL annotation scheme~\cite{core1997coding}. 
The labels also reflect the roles partners take in a conversation. 
The role of the Agent is to provide Answers to User's Queries. During the conversation the Agent may Request additional information from the User and the Agent may provide Feedback to the Agent's actions. 

The initial set of four labels (QRFA) are further subdivided to provide a more fine-grained level of detail.
See Table~\ref{schema-table} and the descriptions below:

\mysubsubsection{Q}uery provides context (or input) for \textit{Information} search (question answering), as the default functionality provided by the agent (e.g., ``Where does ECIR take place this year?''), but can also \textit{Prompt} the agent to perform actions, such as cancel the previous query or request assistance, e.g., ``What options are available?''

\mysubsubsection{R}equest is a pro-active utterance from the agent, when there is a need for additional information (Feedback) from the user. It was the only class that caused disagreement between the annotators, when trying to subdivide it into two groups of requests: the ones that contain an \textit{Offer}, such as an offer to help the User or presenting the options available (e.g., ``I can group the datasets by organization or format''), and the ones whose main goal is to \textit{Understand} the user need, such as requests to repeat or rephrase the Query (e.g., ``Sorry I am a bit confused; please tell me again what you are looking for''). 

\mysubsubsection{F}eedback from the user can be subdivided by sentiment into \textit{Positive}, such as accept or confirm, and \textit{Negative}, such as reject or be discontented.
 
\mysubsubsection{A}nswer corresponds to the response of the agent, which may contain one of the following: (1)~\textit{Results}, such as a search engine result page (SERP) or a link to a dataset, (2)~\textit{Backchannel} response to maintain contact with the User, such as a promise or a confirmation (e.g., ``One moment, I'll look it up.''), and (3)~\textit{Empty} result set (e.g., ``I am sorry but there is no other Indian restaurant in the moderate price range'').

Two authors of this paper independently aligned the annotation schemes of the datasets and the COR model to match the single schema with an inter-annotator agreement of 94\%.
We found the first more abstract level of annotation sufficient for our experiments to make the conversation models easier to interpret. 
The complete table containing alignments across the schemes is made available to the community to enable reproducibility and encourage future work in this direction.%
\footnote{\scriptsize\url{https://github.com/svakulenk0/conversation_mining/blob/master/annotations/alignments_new.xls}}

\subsection{QRFA model dynamics}

We used the ProM Episode Miner plug-in~\cite{DBLP:conf/simpda/LeemansA14} and a declarative process mining tool, MINER\-ful%
\footnote{\scriptsize\url{https://github.com/cdc08x/MINERful}}~\cite{DiCiccio.Mecella/CIDM2013:TwoStepFast,DiCiccio.Mecella/ACMTMIS2015:DiscoveryDeclarativeControl}, to discover frequent sequence patterns in the conversation transcripts.

\begin{figure}[tb]
	\CenterFloatBoxes
	\begin{floatrow}
		\ffigbox{%
			\includegraphics[trim={0 6.5cm 0 0},clip,width=0.45\textwidth]{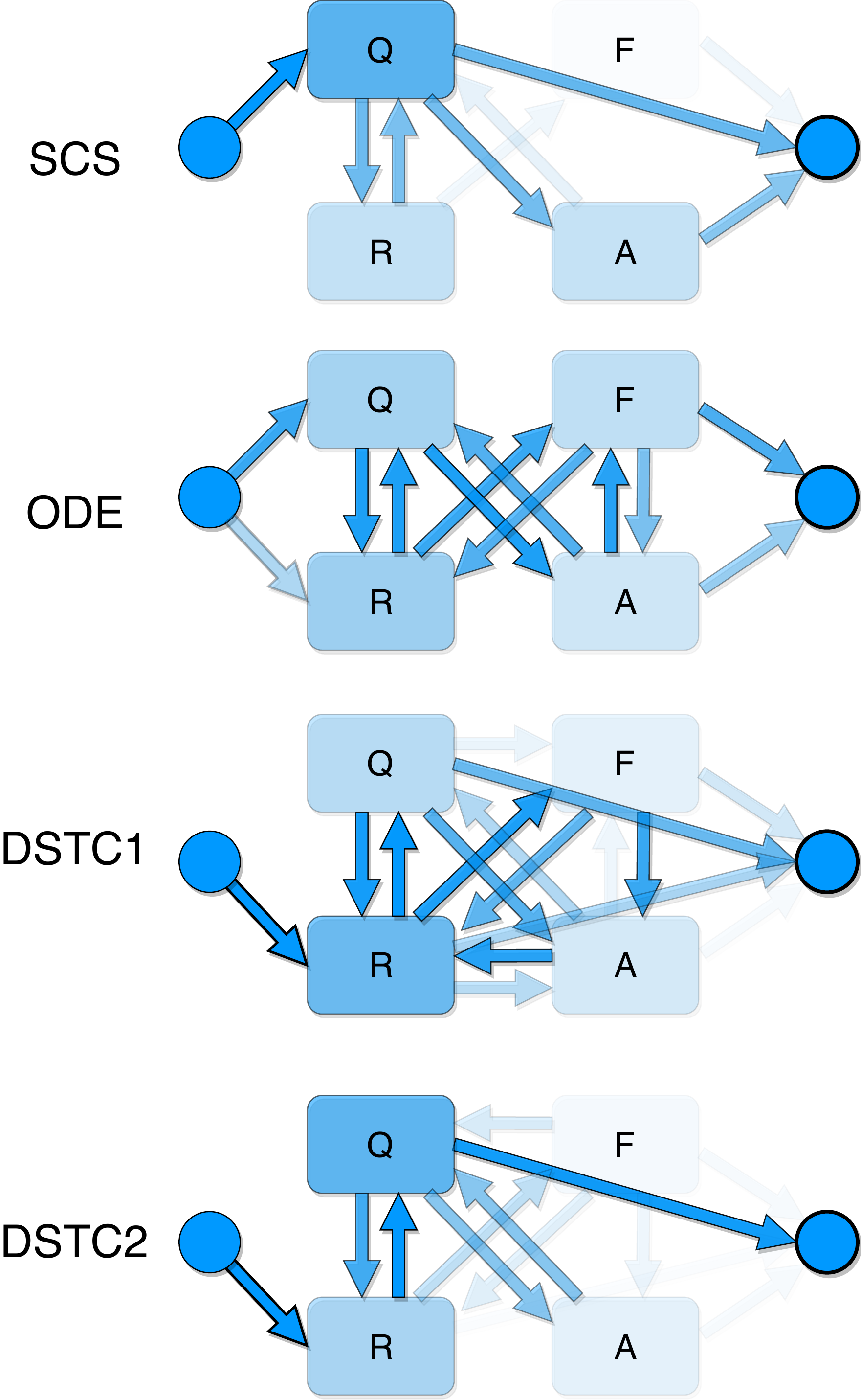} %
		}{%
			\caption{Conversation flows in the SCS, ODE and DSTC1 datasets. Color intensity indicates frequency.} %
			\label{fig:conversation_flows} %
		}
		\killfloatstyle
		\ffigbox{%
			\includegraphics[width=0.45\textwidth]{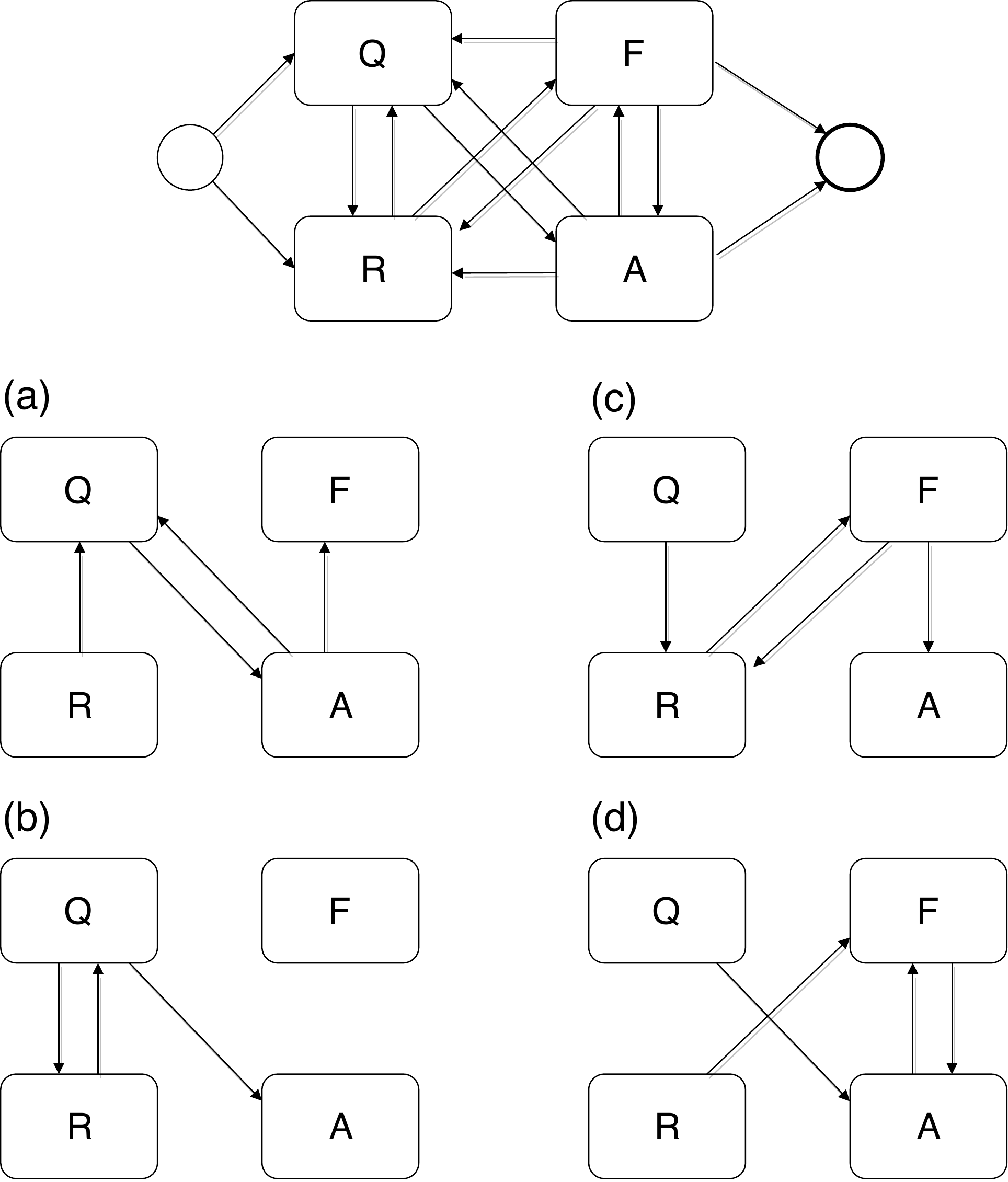} %
		}{%
			\caption{The QRFA model for conversational search composed of the ``four virtuous cycles of information-seeking'': (a) question answering loop, (b) query refinement loop, (c) offer refinement loop, (d) answer refinement loop.} %
			\label{fig:cycles} %
		}
	\end{floatrow}
\end{figure}

Figure~\ref{fig:conversation_flows} illustrates the conversation flows in each of the three datasets used for model discovery (one of the datasets, DSTC2, is held out for model evaluation).
Color intensity (opacity) indicates the frequency of the observed sequences between the pairs of utterances within the respective dataset (the frequency counts for all transitions across all the datasets are available on-line%
\footnote{\scriptsize\url{https://github.com/svakulenk0/conversation_mining/tree/master/results/}}).
An empirically derived information-seeking conversation model would be the sum of the models extracted from the three conversation transcripts.

However, an empirically derived model guarantees neither correctness nor optimality since the transcripts (training data) may contain errors, i.e., negative patterns.
Instead of blindly relying on the empirical ``as is'' model, we analyze and revise it (re-sample) to formulate our theoretical model of a successful information-seeking conversation (Figure~\ref{fig:cycles}).
For example, many conversations in the SCS and DSTC1 datasets are terminated right after the User Query (Q$\rightarrow$END pattern) for an unknown reason, which we consider to be undesirable behavior: the User's question is left unanswered by the Agent.
Therefore, we discard this transition from our prescriptive model, which specifies how a well-structured conversation ``should be'' (Figure~\ref{fig:cycles}).
Analogously to discarding implausible transitions, the power of the theoretical modeling lies in the ability to incorporate transitions are still considered legitimate from a theoretical point of view even though they were not observed in the training examples.
Incompleteness in empirically derived models may stem from assumptions already built into the systems by their designers, when analyzing dialogue system logs, or also differences in the annotation guidelines, e.g., one label per utterance constraint.
In our case, we noticed that adding the FQ transition, which was completely absent from our training examples, will make the model symmetric.
The symmetry along the horizontal axes reflects the distribution of the transitions between the two dialogue partners.
Hence, the FQ transition mirrors the AR transition, which is already present in our transcripts, but on the User side.
The semantics of an FQ transition is that the User can first give feedback to the Agent and then follow up with another question.
\citet[Figure 1, Example 1]{trippas2018informing} empirically show that utterances in information-seeking dialogues tend to contain multiple moves, i.e., can be annotated with multiple labels.

The final shape of a successful information-seeking conversation according to our model is illustrated in Figure~\ref{fig:cycles}.
To analyze this model in more detail, we decompose it into a set of connected components, each containing one of the cycles from the original model. 
We refer to them as ``four virtuous cycles of information-seeking,''\footnote{A virtuous cycle refers to complex chains of events that reinforce themselves through a feedback loop. A virtuous circle has favorable results, while a vicious circle has detrimental results.} representing the possible User-Agent exchanges (feedback loops) in the context of: (a)~question answering, (b)~query refinement, (c)~offer refinement, and (d)~answer refinement. 
To verify that the loops actually occur and estimate their frequencies, we mined up to 4-label sequences from the transcripts using the Episode Miner plug-in.

\subsection{QRFA model evaluation}
Our evaluation of the QRFA model is twofold. 
Firstly, we measure model fitness with respect to the conversational datasets including a held-out dataset (DSTC2), which was not used during model development, to demonstrate the ability of the model to fit well across all available datasets and also generalize to unseen data. 
Secondly, we hypothesize that deviations from the conversation flow captured in the QRFA model signal anomalies, i.e., undesired conversation turns. Therefore, we also compare the model's performance on the task of error detection in conversational transcripts with human judgments of the conversation success.

\mysubsubsection{Fitness and generalization.}
To analyze the model fit with respect to the actual data we applied the conformance checking technique proposed by \citet{AdriansyahDA11EDOC}, available as a ProM plug-in under the name ``Replay a Log on Petri Net for Conformance Analysis.'' 
To this end, we translated the COR and QRFA models into the Petri net notation and ran a conformance analysis for each model on every dataset.
Exit and Restart activities are part of the ``syntactic sugar'' added for the Petri net notation and we set them to \textit{invisible} in order to avoid counting them, when assigning the costs during analysis.

Table~\ref{table:evaluation} (top) contains the fitness measures of the COR and QRFA models for all the datasets. 
We use the default uniform cost function that assigns a cost of~1 to every non-synchronous move. 
Fitness is computed separately for each sequence (dialogue) as a proportion of the correctly aligned events.
We measure generalization as the fitness of a model on the sequences that are not considered for its creation.
The ability of the model to generalize to a different held-out dataset is significant ($0.99$ on average). 
This result demonstrates the out-of-sample generalizability of the model, which is a more challenging task than testing the model on the held-out (test) splits from the same datasets (label frequency distributions) used for the model development.
Remarkably, the baseline COR model managed to fully fit only a single conversation across all four datasets.
This comparison clearly shows the greater flexibility that the QRFA model provides, which in turn indicates the requirement for information-seeking dialogue systems to be able to operate in four different IR modes (Figure~\ref{fig:conversation_flows}) and seamlessly switch between them when appropriate.

\mysubsubsection{Conversation success and error detection.}
Since only one of the datasets, namely ODE, was annotated with a success score, we add manual annotations for the rest of the datasets (2 annotators, inter-annotator agreement: 0.85).
We produced annotations for 89 dialogues in total, for the full SCS dataset and a random sample for each of the DSTC datasets.%
\footnote{\scriptsize{\url{https://github.com/svakulenk0/conversation_mining/tree/master/annotations/dialogue_success}}}
Criteria for the success of a conversational interaction are defined in terms of informational outcomes, i.e., the search results were obtained and the information need was satisfied, as well as emotional outcomes, i.e., whether the interaction was pleasant and efficient.

\begin{table}[t]
\begin{floatrow}
\ttabbox{%
\begin{adjustbox}{width=.98\textwidth}%
\begin{tabular}{lllllllllllllll}
\toprule
\multicolumn{1}{l}{Dataset} & \multicolumn{3}{l}{SCS} & \multicolumn{3}{l}{ODE} & \multicolumn{3}{l}{DSTC1} & \multicolumn{3}{l}{DSTC2} & \multicolumn{2}{l}{Average} 
\\ \cmidrule(r){1-1}\cmidrule(r){2-4}\cmidrule(r){5-7}\cmidrule(r){8-10}\cmidrule(r){11-13}\cmidrule{14-15}
Metric/Model                           & COR      & QRFA     & GS       & COR      & QRFA     & GS       & COR      & QRFA      & GS        & COR      & QRFA      & GS        & COR             & QRFA             \\ \midrule
Average/case                           & 0.58        & 0.89     &          & 0.74        & 1        &          & 0.66        & 0.96      &           & 0.7         & 0.99      &           & 0.67               & 0.96             \\ 
Max.                                   & 0.8         & 1        &          & 1           & 1        &          & 1           & 1         &           & 0.91        & 1         &           & 0.93               & 1                \\ 
Min.                                   & 0.4         & 0.8      &          & 0.6         & 1        &          & 0           & 0         &           & 0.53        & 0.8       &           & 0.38               & 0.65             \\ 
Std. Deviation                         & 0.17        & 0.1      &          & 0.09        & 0        &          & 0.08        & 0.05      &           & 0.05        & 0.02      &           & 0.10               & 0.04             \\ \midrule
Cases with value 1                   & 0           & 0.46     & 0.37     & 0.04        & 1        & 0.92     & 0           & 0.14      & 0.07      & 0           & 0.83      & 0.79      & 0.01               & 0.61             \\ 
Error detection Precision                   &            & 1     &      &         & 1        &      &            &  1    &       &           & 0.67       &       &              &       0.92         \\ 
Error detection Recall                     &            & 0.78     &      &         & 0       &      &            &  0.83    &       &           & 0.57       &       &              &       0.55         \\ 
\bottomrule
\end{tabular}
\end{adjustbox}
}{%
\caption{Evaluation results of QRFA and COR models of information-seeking dialogues on the conversational datasets in terms of model fitness/generalization (top) and error detection abilities (bottom). The gold standard (GS) column refers to the manual annotations of the conversational datasets with the conversation success score (inter-annotator agreement: 0.85).}%
\label{table:evaluation}%
}
\end{floatrow}
\end{table}

Results of the conversation success prediction task are summarized in Table~\ref{table:evaluation} (bottom); QRFA correlates well with human judgments of conversation success based on the model fitness obtained via conformance checking (Cases with value 1).
We also took a closer look at the cases annotated as unsuccessful in terms of fitness to the QRFA model and reported Precision/Recall metrics for the conversation failure detection task.
For example, the model predicted all conversations in the ODE corpus as success (100\% success rate) and overlooked 8\% that actually failed, hence the Recall for conversation failure detection is 0 in this case. 

Table~\ref{table:evaluation} shows that half of the errors affecting conversation success are due to a violation of structural requirements formulated via the QRFA model.
The model overestimates the success rate of a dialog agent since only syntactic information in some cases is not enough to evaluate the overall performance, such as the quality of the answer obtained.
However, it shows very promising results, clearly indicating the faulty cases, such as the situations when the user's query was left unanswered by the agent (SCS and DSTC1).

Our evaluation shows that the QRFA model reflects the patterns of successful information-seeking conversations and the deviations from its shape likely indicate flaws in the conversation flow. 
These results are demonstrated across four conversational datasets from different domains. 
In particular, then, QRFA does not overfit the errors from the datasets used for development and it generalizes to the held-out dataset.

We conclude that the QFRA model satisfies the four quality criteria for a process model defined by \citet{Aalst16}: (1)~fitness -- it fits across four conversational datasets without overfitting, which allows it to successfully detect deviations (errors) in the information-seeking process (Table~\ref{table:evaluation}); (2)~precision -- all types of  interaction described by the model are observed in the conversation transcripts; (3)~generalization -- the model is able to describe the structure and deviations in previously unseen conversations; and (4)~simplicity -- it contains a minimal number of elements necessary to describe the conversation dynamics in information-seeking dialogues.


\section{Conclusion}
\label{sec:conclusion}

We have proposed an annotation schema and a theoretical model of information-seeking dialogues grounded in empirical evidence from several public conversational datasets.
Our annotation schema resembles the approach used in DAMSL, where utterances are classified into Forward and Backward Communicative Function, but adopts labels to our information-seeking setting, where roles are more distinct due to information asymmetry between participants.
The patterns that we have discovered extend and correct the assumptions built into the COR model and also incorporate frameworks previously proposed within the information retrieval community.

Our empirical evaluation indicates that, however simple, the QRFA model still provides a better fit than the most comprehensive model proposed previously by explaining the conversational flow in the available information-seeking conversational datasets. 
Moreover, we have described an efficient way to provide sanity checking diagnostics of a dialogue system using process mining techniques (conformance checking) and have shown how the QRFA model helps to evaluate the performance of an existing dialogue system from its transcripts. 

In future work, we plan to evaluate the QRFA model against new conversational datasets and further extend it to a finer granularity level if required.
Our experiments so far have utilized hand-labeled conversation transcripts.
Introducing automatically generated labels may propagate errors into the model extraction phase.
Nevertheless, discovering patterns in raw conversational data that is automatically tagged with semantic labels is an exciting research direction~\cite{DBLP:conf/semweb/VakulenkoRCSP18}.
In addition, the predictions of the QRFA model may be an informative signal for evaluating or training reinforcement learning-based dialogue systems~\cite{li-dialogue-2019}.

Wide adoption of information-seeking dialogue systems will lead to a massive increase in conversational data, which can potentially be used for improving dialogue systems. 
We believe that QRFA and similar models will become important for informing the design of dialogue systems, motivating collection of new information-seeking conversational data, specifying the functional requirements the systems should satisfy, and providing means for their evaluation.

\subsubsection*{Acknowledgements.}
The work of S.\ Vakulenko and C.\ Di Ciccio has received funding from the EU H2020 program under MSCA-RISE agreement 645751 ({RISE\_BPM}) and the Austrian Research Promotion Agency (FFG) under grant 861213 (CitySPIN).
S.\ Vakulenko was also supported by project 855407 ``Open Data for Local Communities'' 
(CommuniData) of the Austrian Federal Ministry of Transport, Innovation and Technology (BMVIT) under the program ``ICT of the Future.''
M.\ de Rijke was supported by Ahold Delhaize, the Association of Universities in the Netherlands (VSNU), and the Innovation Center for Artificial Intelligence (ICAI).

All content represents the opinion of the authors, which is not necessarily shared or endorsed by their respective employers and/or sponsors.

\bibliographystyle{abbrvnat}
\bibliography{bibliography/bib,bibliography/CDC-IS2018}

\end{document}